\documentclass[format=sigconf, anonymous=false, screen=true]{acmart}

\copyrightyear{2021}
\acmYear{2021}
\setcopyright{acmlicensed}
\acmConference[SIGIR '21]{Proceedings of the 44th International ACM SIGIR Conference on Research and Development in Information Retrieval}{July 11--15, 2021}{Virtual Event, Canada}
\acmBooktitle{Proceedings of the 44th International ACM SIGIR Conference on Research and Development in Information Retrieval (SIGIR '21), July 11--15, 2021, Virtual Event, Canada}
\acmPrice{15.00}
\acmDOI{10.1145/3404835.3462849}
\acmISBN{978-1-4503-8037-9/21/07}

\newtheorem{thm}{Theorem}

\newcommand{\return}{\vspace{0.8mm}}
\newcommand{\parag}[1]{\vspace{1.5mm}\noindent\textbf{{#1.}}}

\newcommand{\aftertabspace}{\vspace{-2.5mm}}

\newcommand{\mytablescale}{0.9} 
\newcommand{\mycaption}[1]{\caption{{\rm{#1}}}}

\newcommand{\tabtimes}[1]{\small(\ensuremath{#1\times})}

\newcommand{\mymod}[1]{\,\textup{mod}{\,#1}}
\newcommand{\code}[1]{\mbox{\textsf{#1}}}
\newcommand{\var}[1]{\mbox{\emph{#1}}}
\newcommand{\bit}[1]{\mbox{\texttt{#1}}}
\newcommand{\func}[1]{\mbox{\emph{#1}}}

\usepackage{xcolor, soul}
\usepackage[normalem]{ulem} 
\newcommand{\remove}[1]{\bgroup\markoverwith{\textcolor{red}{\rule[1pt]{1pt}{2pt}}}\ULon{#1}}

\newcommand{\nspk}{{ns/key}}
\newcommand{\bpk}{{bits/key}}

\usepackage{pbox}
\usepackage{makecell}
\usepackage{tikz}
\usetikzlibrary{calc}

\renewcommand{\O}{\mathcal{O}}

\newcommand{\method}[1]{\textup{{#1}}}
\newcommand{\pth}{\method{PTHash}}

\definecolor{tomato}{HTML}{E74C3C}
\definecolor{gray}{HTML}{D5D8DC}
\definecolor{darkgray}{HTML}{707070}
\definecolor{green}{HTML}{D5F5E3}
\usepackage{graphicx}
\usepackage{balance}

\usepackage{colortbl}
\usepackage{xcolor}

\usepackage{bm}
\usepackage{url}
\usepackage{amsmath}
\usepackage{multirow}
\usepackage{booktabs}
\usepackage{makecell}
\usepackage[caption=false]{subfig}
\usepackage{wrapfig}
\usepackage{xcolor, soul}
\usepackage{siunitx}
\usepackage{enumitem}
\usepackage{hyperref}
\usepackage{rotating}
\usepackage{tablefootnote}
\usepackage[noend]{algpseudocode}
\usepackage[linesnumbered
			,vlined]{algorithm2e}
\usepackage{listings}
\setlist[itemize,1]{leftmargin=3mm,itemsep=0.8mm}
\setlist[enumerate,1]{leftmargin=3mm,itemsep=0.8mm}

\begin{document}

\fancyhead{}

\title{PTHash: Revisiting FCH Minimal Perfect Hashing}

\author{Giulio Ermanno Pibiri}
\affiliation{%
  \institution{ISTI-CNR, Pisa, Italy}
}
\email{giulio.ermanno.pibiri@isti.cnr.it}

\author{Roberto Trani}
\affiliation{%
  \institution{ISTI-CNR, Pisa, Italy}
}
\email{roberto.trani@isti.cnr.it}


\begin{abstract}
Given a set $S$ of $n$ distinct keys,
a function $f$ that bijectively maps the keys
of $S$ into the range $\{0,\ldots,n-1\}$
is called a \emph{minimal perfect hash function} for $S$.
Algorithms that find such functions when $n$ is large
and retain \emph{constant} evaluation time are of practical interest;
for instance, search engines
and databases typically use
minimal perfect hash functions to quickly assign identifiers
to static sets of variable-length keys such as strings.
The challenge is to design an algorithm
which is \emph{efficient} in three different aspects:
time to find $f$ (construction time),
time to evaluate $f$ on a key of $S$ (lookup time),
and space of representation for $f$.
Several algorithms have been proposed to trade-off
between these aspects.
In 1992, Fox, Chen, and Heath (FCH) presented
an algorithm at SIGIR
providing very fast lookup evaluation.
However,
the approach received little attention because
of its large construction time and higher space consumption
compared to other subsequent techniques.
Almost thirty years later we revisit their framework
and present an improved algorithm
that scales well to large sets and
reduces space consumption altogether,
without compromising the lookup time.
We conduct an extensive experimental assessment and
show that the algorithm finds functions that
are competitive in space with
state-of-the art techniques and provide $2-4\times$ better lookup time.
\end{abstract}



\maketitle



\section{Introduction}\label{sec:introduction}

The \emph{minimal perfect hashing} problem is to
build a data structure that assigns the numbers
in $[n]=\{0,\ldots,n-1\}$ to the $n$ distinct
elements of a static set $S$.
The resulting data structure is called a
minimal perfect hash function $f$, or MPHF in short,
and should consume little space while supporting
constant-time evaluation for any key in $S$.
In principle, it would be trivial to obtain a MPHF
for $S$ using
a perfect hash table: each key $x \in S$
is associated in $\O(1)$ to the pair $\langle x,p \rangle$
stored in the table,
where $p$ is the ``identifier'' of $x$ in $[n]$.
However, this approach pays the cost of
representation for the set $S$ itself, in addition to the
storage cost of the numbers in $[n]$, which occupy
$\Theta(n \log n)$ bits.
Nevertheless, the minimal perfect hashing problem ignores the
behaviour of $f$ on keys that are \emph{not} in $S$.
This relaxation makes it possible to discard the space
for $S$, thus admitting a space lower bound of
$\log_2 e \approx 1.44$
{\bpk}~\cite{fredman1984storing,mehlhorn1982program}
regardless the size and type of the input.


Practical applications of minimal perfect hashing
are pervasive in computing and involve
compressed full-text indexes~\cite{belazzougui2014alphabet},
computer networks~\cite{lu2006perfect},
databases~\cite{chang2005perfect},
prefix-search data structures~\cite{belazzougui2010fast},
language models~\cite{pibiri2017efficient,PibiriV19},
Bloom filters and variants~\cite{broder2004network,fan2014cuckoo,graf2020xor},
just to name a few.

Several algorithms are known for this problem,
each of them exposing a trade-off between
construction time, lookup time, and space consumption.
We review these approaches in Section~\ref{sec:related_work}.
Our starting point for this work
is an old technique, originally
described by~\citet*{fox1992faster}, and named FCH after them.
They proposed a three-step framework to build a data structure that
allows to evaluate the MPHF with just a single
memory access to an array, besides a few hash and arithmetic calculations.
More specifically, the algorithm builds a MPHF that occupies
$cn$ bits for a parameter $c$.
This parameter drives the efficiency of the construction:
the higher the value of $c$, the lower the time to construct $f$.
Unfortunately, to find a MPHF with FCH in a reasonable amount of time
on large inputs, one has to use a large value of $c$
and, thus, waste space.
On the other hand, more recent algorithms scale up very well
and take less {\bpk},
but usually perform 3-4 memory accesses per lookup.
For this reason, this work aims
at combining the fast lookup time of FCH with
a more compact representation.

It should be noted that, once the space of the MPHF is \emph{sufficiently
small} albeit not very close to the theoretic minimum,
the most important aspect becomes indeed lookup time
\emph{provided that} the function can be built feasibly~\cite{limasset2017fast}.
In fact, in practical applications,
MPHFs are employed as
building blocks to obtain static dictionaries,
i.e., key-value stores, which are built \emph{once} and evaluated many times.
If the values take much more space
than the MPHF itself, which is the typical case in practice,
whether the MPHF takes 3 or 2 bits per key
is not a critical issue.
Fast lookup and feasible construction time are, therefore,
the most critical aspects for this problem,
hence the ones we focus on.

In this paper, we propose {\pth} --- a novel algorithm combining the lookup
performance of FCH with succinct space and
fast construction on large sets.
The crucial aspect of {\pth} is that it
dramatically reduces the entropy of the information
stored in the data structure compared to FCH.
This, in turn, permits to tailor a light-weight compression
scheme that reduces space consumption significantly
while maintaining noticeable lookup performance.


We evaluate {\pth} on large sets of fixed and variable-length keys,
in comparison to state-of-the-art techniques.
We show that it is competitive in construction
time with other techniques and,
for approximately the same space consumption,
it is $2-4\times$ faster at lookup time.
Our C++ implementation of {\pth} is
available at
\url{https://github.com/jermp/pthash}.



\section{Related Work}\label{sec:related_work}
Up to date, four different approaches have been devised
to solve the minimal perfect hashing problem.
We summarize them in chronological order of proposal.
We also point out that some theoretical constructions,
like that by~\citet{hagerup2001efficient},
can be proved to reach the space lower bound of $n\log_2 e$ bits,
but only work in the asymptotic sense, i.e., for $n$
too large to be of any practical interest.

\parag{Hash and Displace}
The ``hash and displace'' technique was originally introduced
by~\citet*{fox1992faster} (although \citet*{pagh1999hash}
named it this way inspired by a work due to~\citet{tarjan1979storing}).
Since we describe their approach (FCH) in details in Section~\ref{sec:FCH},
we just give a simple overview here.
Keys are first hashed and mapped into \emph{non-uniform} buckets $\{B_i\}$;
then, the buckets are sorted and processed by falling size:
for each bucket $B_i$, a displacement value $d_i \in [n]$ is determined
so that all keys in the bucket can be placed without
collisions to positions $(h(x) + d_i) \mymod n$,
for a proper hash function $h$ and $x \in B_i$.
Lastly, the sequence of displacements $d_i$ is stored in compact form
using $\lceil \log_2 n \rceil$ bits per value.
While the theoretical analysis suggests that by
decreasing the number of buckets
it is possible to lower the space usage
(at the cost of a larger construction time),
in practice it is unfeasible to go below $2.5$ {\bpk} for large values of $n$.

In the compressed hash and displace (CHD) variant by~\citet{belazzougui2009hash},
keys are first \emph{uniformly} distributed to buckets,
with expected size $\lambda > 0$;
then, for each bucket $B_i$,
a pair of displacements $\langle d_0,d_1 \rangle$ is determined
so that all keys in the bucket can be placed without
collisions to positions
$(h_1(x) + d_0 h_2(x) + d_1) \mymod n$,
for a given pair of hash functions $h_1, h_2$ and for $x \in B_i$.
Instead of explicitly storing a pair $\langle d_0,d_1 \rangle$
for each bucket, the index of such pair in the sequence
$$\langle 0,0 \rangle,...,\langle 0,n-1 \rangle,\langle 1,0 \rangle,...,\langle 1,n-1 \rangle,...,\langle n-1,0 \rangle,...,\langle n-1,n-1 \rangle$$
is stored.
Lastly, the sequence of indexes is stored in compressed form
using the entropy coding mechanism introduced by~\citet*{fredriksson2007simple},
retaining $\O(1)$ access time.

\parag{Linear Systems}
In the late $90$s, \citet{majewski1996family} introduced an algorithm
to build a MPHF exploiting a connection between
linear systems and hypergraphs.
(Almost ten years later, \citet{chazelle2004bloomier} proposed an analogous
construction in an independent manner.)
The MPHF $f$ is found by generating a system of $n$ random equations in $m$
variables of the form
$$w_{h_1(x)} + w_{h_2(x)} + \cdots + w_{h_r(x)} = f(x) \mymod{n}, x \in S,$$
where $h_i : S \rightarrow [m]$ is a random hash function,
and $\{w_i\}$ are $m$ variables whose values are in $[n]$.
Due to bounds on the acyclicity of random graphs,
if the ratio between $m$ and $n$ is above a certain
threshold $\gamma_{r}$, the system can be almost always triangulated and solved
in linear time by peeling the corresponding hypergraph.
The constant $\gamma_{r}$ depends on the degree $r$ of the graph,
and attains its minimum for
$r=3$, i.e., $\gamma_{3} \approx 1.23$.

\citet{belazzougui2014cache} proposed a cache-oblivious implementation
of the algorithm suitable for external memory constructions.
\citet{genuzio2016fast,genuzio2020fast} demonstrated practical
improvements to the Gaussian elimination technique used to solve the linear
system, that make the overall construction, lookup time, and space competitive
with the cache-oblivious implementation~\cite{belazzougui2014cache}.

\parag{Fingerprinting}
\citet{muller2014retrieval} introduced
a technique based on fingerprinting.
The general idea is as follows.
All keys are first hashed in $[n]$ using a random hash function
and collisions are recorded using a bitmap $B_0$ of size $n_0=n$.
In particular,
keys that do not collide have their position
in the bitmap marked with a \bit{1};
all positions involved in a collision are marked with
a \bit{0} instead. If $n_1 > 0$ collisions are produced,
then the
same process is repeated recursively for the $n_1$
colliding keys using a bitmap $B_1$ of size $n_1$.
All bitmaps, called ``levels'', are then concatenated together
in a bitmap $B$.
The lookup algorithm keeps hashing the key level by level
until a \bit{1} is hit, say in position $p$ of $B$.
A constant-time ranking data structure~\cite{jacobson1989space,navarro2016compact}
is used to count the number of \bit{1}s in $B[0..p]$
to ensure that the returned value is in $[n]$.
On average, only $1.56$ levels are accessed
in the most succinct setting~\cite{muller2014retrieval} (3 {\bpk}).


\citet{limasset2017fast} provided an implementation
of this approach, named BBHash, that is very fast in construction
and lookup, and scales to very large key sets using multiple threads.
A parameter $\gamma \geq 1$ is introduced to speedup
construction and query time, so that bitmap $B_i$
on level $i$
is $\gamma n_i$ bits large.
This clearly reduces collisions and, thus, the average
number of levels accessed at query time.
However, the larger $\gamma$, the higher the space consumption.

\parag{Recursive Splitting}
Very recently, \citet{esposito2020recsplit} proposed
a new technique, named RecSplit,
for building very succinct MPHFs in expected
linear time and providing expected constant lookup time.
The authors first observed that for very small sets
it is possible to find a MPHF simply by brute-force searching
for a bijection with suitable codomain.
Then, the same approach is applied in a divide-and-conquer manner
to solve the problem on larger inputs.

Given two parameters $b$ and $\ell$, the keys are divided into buckets
of average size $b$ using a random hash function.
Each bucket is recursively
split until a block of size $\ell$ is formed
and for which a MPHF can be found using brute-force, hence forming
a rooted tree of splittings.
The parameters $b$ and $\ell$ provide different space/time trade-offs.
While providing a very compact representation,
the evaluation performs one memory access for each level of the tree
that penalizes the lookup time.


\section{The FCH Technique}\label{sec:FCH}
Fox, Chen, and Heath presented a three-step framework~\cite{fox1992faster}
for minimal perfect hashing in 1992, which we describe here as it
will form the basis for our own development in Section~\ref{sec:main}.
For a given set $S$ of $n$ keys,
the technique finds a MPHF $f$ of size $cn$ bits, where
$c > \log_2 e$ is a parameter that
trades space for construction time.

\parag{Construction}
The construction is carried out in three steps,
namely \emph{mapping}, \emph{ordering}, and \emph{searching}.
First, the mapping step partitions the set of keys by placing
the keys into \emph{non-uniform} buckets.
Then, the ordering step orders the buckets by \emph{non-increasing} size,
which is the order used to process the keys.
Lastly, the searching step attempts to assign
the free positions in $[n]$
to the keys of each bucket, in a greedy way.
Two hash functions are used during the process.
In practice, a pseudo random function $h$ is used,
such as MurmurHash~\cite{smhasher},
with two different seeds $s_1$ and $s_2$.
The seed $s_1$ can be chosen at random,
whereas $s_2$ must satisfy a property as explained shortly.


\return\noindent(i) \emph{Mapping.}
The non-uniform mapping of keys into buckets is accomplished
as follows.
All keys are hashed using $h(\cdot,s_1)$
and a value $p_1$ is chosen such that $S$
is partitioned into two sets, $S_1 = \{x | (h(x,s_1)\mymod{n}) < p_1\}$
and $S_2 = S / S_1$.
Then, $m = \lceil cn / (\log_2 n + 1) \rceil$ buckets are
allocated to hold the keys, for a given $c$.
The lower the value of $c$ the lower the number of buckets and
the higher the average number of keys per bucket.
Finally, a value $p_2$ is chosen so that
each key $x \in S$ is assigned to the bucket

\begin{equation}\label{eq:bucket}
\func{bucket}(x) =
   \left\{
\begin{array}{ll}
      h(x,s_1)\mymod{p_2} & x \in S_1 \\
      p_2 + \left(h(x,s_1) \mymod{(m-p_2)}\right) & \mbox{otherwise} \\
\end{array}
\right..
\end{equation}

\noindent
The arbitrary thresholds $p_1$ and $p_2$ are set to,
respectively, $0.6n$ and $0.3m$ so that the mapping
of keys into the buckets is \emph{uneven}:
roughly 60\% of the keys are mapped to the 30\% of the
buckets.

\return\noindent(ii) \emph{Ordering.}
After all keys are mapped to the $m$ buckets,
the buckets are sorted by \emph{non-increasing} size to speed up the searching step.

\return\noindent(iii) \emph{Searching.}
For each bucket $B_i$ in the order given by the previous step,
a \emph{displacement} value $d_i$
and an extra bit $b_i$ are determined so that
all keys in the bucket do not produce collisions
with previously occupied positions,
when hashed to positions
$$
\func{position}(x,d_i,b_i) = (h(x,s_2+b_i) + d_i)\mymod{n}.
$$
In particular, the global seed $s_2$ is chosen such that
there are no collisions between
the keys of the \emph{same} bucket.
To add a degree of freedom when searching,
an extra bit $b_i$ is used for the seed of $h$
for bucket $B_i$ as to avoid failures
when no displacement is found:
after trying all displacements for $b_i=\bit{0}$,
the same displacements are tried again with $b_i=\bit{1}$.
To accelerate the identification of $d_i$, an auxiliary data structure
formed by two integer arrays of $\Theta(n \log n)$ bits is used
to locate the available positions.
Given a random available position $p$,
$d_i$ is computed ``aligning'' an arbitrary key of $B_i$ to $p$.
This random alignment is rather critical for the algorithm as it guarantees
that all positions have the \emph{same} probability to be occupied,
thus, preserving the search efficiency.

\parag{Discussion}
When the searching phase terminates,
the $m$ pairs $\langle d_i, b_i \rangle$, one for each bucket,
have been determined and stored in an array $P$,
in the order given by the function \func{bucket}.
This array is what the MPHF actually stores.
Since at most $n$ displacement values can be tried
for a bucket, each $d_i$ needs $\lceil \log_2 n \rceil$ bits.
Therefore, as $\lceil\log_2 n\rceil + 1$ bits are spent
to encode the pair $\langle d_i, b_i \rangle$,
the total space for storing the array $P$
is $m (\lceil\log_2 n\rceil + 1) \approx cn$ bits.

With the array $P$, evaluating $f(x)$ is simple
as shown in the following pseudocode.
Although the original authors did not discuss
implementation details, a \emph{single} memory
access is needed by the lookup algorithm if
we interleave the two quantities $d_i$ and $b_i$ in a contiguous
segment of $\lceil\log_2 n\rceil + 1$ bits.
In particular, $b_i$ is written as the least significant bit
(from the right) and $d_i$ takes the remaining $\lceil\log_2 n\rceil$
most significant bits. To recover the two quantities from the
interleaved value $v$ read from $P$, simple and fast bitwise operations are employed:
$b_i = v \,\code{\&}\, \bit{1}$; $d_i = v \gg \bit{1}$.
This is the actual implementation of the assignment
$\langle d_i, b_i \rangle = P[i]$ in step 3
of the pseudocode.
Therefore, the FCH technique achieves very fast
lookup time.

\vspace{-4mm}
\SetArgSty{textnormal}
\begin{algorithm}
\SetKwBlock{Begin}{}{}
\Begin({$f$}\text{(}$x$\text{)} :)
{
    $i = \func{bucket}(x)$ \\
    $\langle d_i, b_i \rangle = P[i]$ \\
    \code{return} $\func{position}(x,d_i,b_i)$
}
\end{algorithm}
\vspace{-4mm}

The constant $c$, fixed before construction,
governs a clear trade-off between space and construction time:
the larger, the higher the probability of finding displacements
quickly because fewer keys per buckets
have to satisfy the search condition simultaneously.
On the other hand, if $c$ is chosen to be small, e.g., $c=2$,
space consumption is reduced but
the searching for a good displacement value may last
for too long --- up to the point when \emph{all} possible $n$
displacements have been tried (for both choices of $b_i$)
and construction fails: at that
point, one has to re-seed the hash functions and try again.

In practice, FCH spends a lot of time to find
a MPHF compared to other more recent techniques
for the \emph{same} space budget.
To make a concrete example,
the algorithm takes 1 hour and 10 minutes to build a MPHF with $c=3$
over a set of 100 million random 64-bit keys.
Other techniques reviewed in Section~\ref{sec:related_work}
are able to do the same in few minutes, or even less,
for the \emph{same} final space budget of 3 {\bpk}.
We will present a more detailed comparison
in Section~\ref{sec:overall_comparison}.



\section{PTH\lowercase{ash}}\label{sec:main}

Inspired by the framework discussed in Section~\ref{sec:FCH},
we now present {\pth},
an algorithm that aims at preserving the lookup
performance of FCH while improving construction
and space consumption altogether.
To do so, we maintain the first two steps
of the framework, i.e., mapping and ordering,
but re-design completely the most critical one
--- the searching step. The novel search algorithm
allows us to achieve some important goals that
are introduced in the following.

Before illustrating the details,
we briefly sketch the intuition behind {\pth}.
Recall that FCH finds a MPHF whose size is $cn$ bits, for a given $c$.
However, FCH
does not offer opportunities for compression:
for each bucket the displacement is chosen with uniform probability from $[n]$
to guarantee that all positions are occupied with equal probability during the search.
This makes the output of the search --- the array $P$ ---
hardly compressible so that $\lceil \log_2 n \rceil$ bits per
displacement is the best we can hope for.
Now, if $P$ were compressible instead,
we could afford to use a different constant $c^{\prime} > c$ for searching,
such that the size of the
\emph{compressed} $P$ would be approximately $cn$ bits.
That is, for the \emph{same space} budget, we search for a MPHF
with a larger $c$, hence reducing construction time.
Our refined ambition in this section is, therefore,
to achieve two related goals:

\return
\noindent(i) design an algorithm that guarantees a compressible output;\\
\noindent(ii) introduce an encoding scheme that does not compromise lookup performance.

\subsection{Searching}
We keep track of
occupied positions in $[n]=\{0,\ldots,n-1\}$ using a bitmap, $\var{taken}[0..n-1]$.
This will be the main supporting structure for
the search and costs just $n$ bits
(we also maintain a small integer vector
to detect in-bucket collisions, but its space is negligible
compared to $n$ bits).
We map keys into $m = \lceil cn / \log_2 n \rceil$ buckets,
$B_0,\ldots,B_{m-1}$,
using Formula (\ref{eq:bucket})
and process them in order of non-increasing size.
We choose a seed $s$ for a pseudo-random hash function $h$
so that keys in the same bucket are hashed to distinct
hash codes.

Now, for each bucket $B_i$, we search
for an integer $k_i$
such that the position assigned to $x \in B_i$ is
\begin{equation}
\func{position}(x,k_i) = (h(x,s) \oplus h(k_i,s)) \mymod{n}
\end{equation}
and
\begin{equation}\label{search_condition}
\var{taken}[\func{position}(x,k_i)] = \bit{0}.
\end{equation}
The quantity $(x \oplus y)$ represents the bitwise XOR between
$x$ and $y$.
If the search for $k_i$ is successful, i.e. $k_i$ places all keys of the bucket $B_i$
into unused positions,
then $k_i$ is saved in the array $P$ and the positions are marked as occupied
via $\var{taken}[\func{position}(x,k_i)]=\bit{1}$;
otherwise, a new integer $k_i$ is tried.

We call the integer $k_i$ a \emph{pilot} for bucket $B_i$
because it uniquely defines the positions of the keys in $B_i$,
hence $P$ is a \emph{pilots table} (PT).

Differently from FCH,
now the random ``displacement'' of keys happens
by virtue of the bitwise XOR instead of using its auxiliary data structure.
More specifically, $\func{position}(x,k_i)$
is based on the principle that the XOR between two random integers
is another random integer.
If fact, since both integers involved in the XOR are
produced by a random hash function we expect them to
have approximately the same proportion of \bit{1} and \bit{0}
bits in their (fixed-size) binary representation.
The bitwise XOR of two such integer is another integer
where the proportion is preserved.

Note that the way $\func{position}(x,k_i)$ is computed assumes
that $n$ is \emph{not} a power of 2 so that all bits
resulting from the XOR are relevant for the modulo.
If $n$ is a power of 2, we just map keys to $[n+1]$
and manage the extra position as we are
going to see in Section~\ref{sec:load_factor}.

This new searching strategy has some direct implications.
First, the XOR between $h(x,s)$ and $h(k_i,s)$ induces a
random ``displacement'' of the keys in $B_i$ where only
the second term changes when changing $k_i$,
allowing thus to precompute the hashes of all keys.
Avoiding multiple calculations of $h(x,s)$
is very important when working with long
keys such as strings.
Second, as we already noted, the space of the auxiliary data structure is spared,
which improves the memory consumption during construction
compared to FCH.
Third, the resilience of the algorithm to failures is improved,
because there is theoretically no limit to the
number of different pilots $k_i$ that can be tried for a bucket.
The displacement values used by FCH, instead,
can only assume values in $[n]$ and
using one extra bit per bucket makes the number of trials
to be $2n$, at most.

But there is also another, very important, implication:
pilots $k_i$ do \emph{not} need to be tried at random,
like displacements values of FCH.
Therefore, our strategy is to choose $k_i$ as the \emph{first} value
of the sequence
$K = 0,1,2,3,\ldots$
that satisfies the search
condition in (\ref{search_condition}),
with the underlying principle being that
$\func{position}(x,k_i)$ does \emph{not} look ``more random''
if also $k_i$ is tried at random.
We argue that the combination of these two effects ---
(i) random positions generated with every $k_i$
and (ii) smaller values always tried first ---
makes $P$ compressible.

To understand why $P$ is now compressible, let us derive a formula for
the expected pilot value $\mathbb{E}[k_i]$.
Each $k_i$ is a random variable taking value $v \in K$
with probability depending on the
current load factor of the bitmap \var{taken}
(fraction of positions marked with \bit{1}).
It follows that $k_i$ is \emph{geometrically distributed}
with success probability $p_i$
being the probability of placing all keys
in $B_i$ without collisions.
Let $\alpha(i)$ be the load factor of the bitmap
after buckets $B_0,\ldots,B_{i-1}$ have been processed, that is
$$
\alpha(i) = \frac{1}{n}\sum_{j=0}^{i-1}|B_j|, \text{ for } i = 1,\ldots,m-1
$$
and $\alpha(0)=0$ for convenience (empty bitmap).
Then the probability $p_i$ can be modeled\footnote{We actually assume that there are no collisions
among the keys in the same
bucket, which is reasonable since $|B_i| \ll n$.
In fact, their influence is negligible in practice.}
as
$
p_i = (1 - \alpha(i))^{|B_i|}.
$
Since $k_i$ is geometrically distributed,
the probability that $k_i=v$, corresponding to
the probability of having success after $v$ failures
($v+1$ total trials),
is $p_i(1-p_i)^v$, with expected value
\begin{equation}\label{eq:expectation}
\mathbb{E}[k_i] = \frac{1}{p_i} - 1 = \Big(\frac{1}{1 - \alpha(i)}\Big)^{|B_i|} - 1.
\end{equation}

\begin{figure}[t]
\centering

\hspace{14mm}\includegraphics[scale=0.8]{{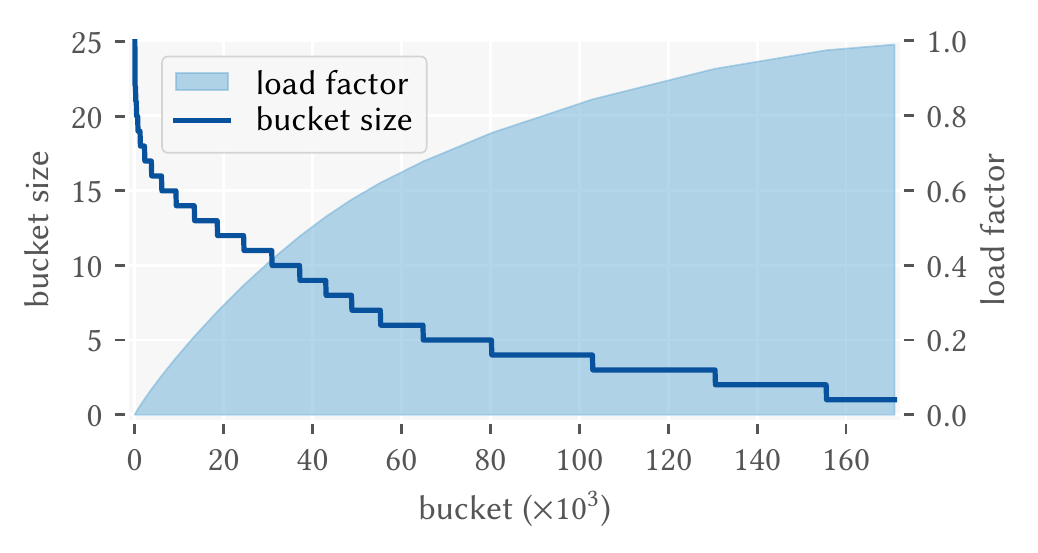}}
\vspace{-9mm}
\mycaption{Bucket size and load factor,
for $n=10^6$ keys and $c=3.5$.
\label{fig:bucket_size}}

\hspace{-10.4mm}\includegraphics[scale=0.8]{{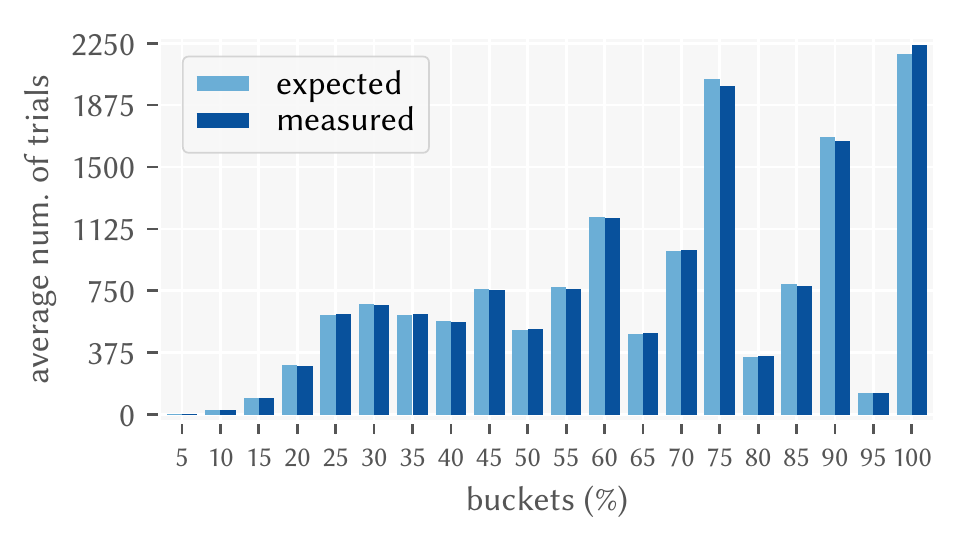}}
\vspace{-5mm}
\mycaption{Average number of trials as buckets are processed
with a step of 5\%,
for $n=10^6$ keys and $c=3.5$.
\label{fig:trials}}
\vspace{-1mm}
\end{figure}

Formula (\ref{eq:expectation}) shows that pilots $k_i$ are quite small, on average.
In fact, $\alpha(i)$ will be
very small for the first processed buckets, hence yielding small $k_i$s.
On the other hand, the ordering of buckets by falling size
plays a crucial role in
keeping $k_i$s small even for high load factors.
In fact, since the $n$ keys are divided into
$m = \lceil cn/\log_2 n \rceil$ non-uniform buckets,
$|B_i|$ is a small quantity between 0
and something larger than the average load $n/m$
i.e., $0 \leq |B_i| \leq \Theta(\frac{\log n}{c})$.
Figure~\ref{fig:bucket_size} shows a pictorial example
of such falling distribution of bucket size,
for $n=10^6$ and $c=3.5$,
with growing load factor as buckets are processed by the search.

We argue that Formula (\ref{eq:expectation})
gives accurate estimates of the average value of $k_i$,
hence of the average number of trials needed by the search.
Figure~\ref{fig:trials} shows the average number of trials,
both measured and expected using Formula (\ref{eq:expectation}),
for $n=10^6$ keys and $c=3.5$.
As evident, the expectation is almost perfectly equal
to the measured value for all buckets.
The second important thing to note is that
while the number of trials tends to
increase as buckets are processed for the growing load factor,
it can still be very small
as it happens, for example, in correspondence of 65\%, 80\% or 95\%.
This is a consequence of the small exponent $|B_i|$
in the formula.

Directly from Formula (\ref{eq:expectation}), we obtain the
following result which relates the performance
of the search with the parameter $c$.
%
%
\begin{thm}\label{thm:expected_runtime}
The expected time of the search,
for $n$ keys and a parameter $c > \log_2 e$,
is
$\O\big(n^{ 1 + \Theta(1/c)}\big).$
\end{thm}
(Proof omitted due to space constraints.)

\begin{table}[t]
\centering
\mycaption{Empirical entropy of $P$
for $n=10^6$ keys, by varying $c$.
\label{tab:entropy}}
\aftertabspace
\scalebox{\mytablescale}{\setlength{\tabcolsep}{2.1pt}
\begin{tabular}{l rrrrrrrrrr}
\toprule
$c \rightarrow$ & 2.5 & 3.0 & 3.5 & 4.0 & 4.5 & 5.0 & 5.5 & 6.0 & 6.5 & 7.0 \\

\midrule

FCH
& 16.69 & 16.85 & 16.93 & 16.93 & 16.87 & 16.77 & 16.65 & 16.49 & 16.32 & 16.14 \\

{\pth}
& 13.42 & 11.68 & 10.32 & 9.29 & 8.48 & 7.82 & 7.27 & 6.82 & 6.45 & 6.11 \\

\bottomrule
\end{tabular}
}
\end{table}

\begin{table}[t]
\centering
\mycaption{Empirical entropy of the \var{front} and \var{back}
parts of $P$
for $n=10^6$ keys, by varying $c$.
\label{tab:entropy_front_back}}
\aftertabspace
\scalebox{\mytablescale}{\setlength{\tabcolsep}{4pt}
\begin{tabular}{l rrrrrrrrrr}
\toprule
$c \rightarrow$ & 2.5 & 3.0 & 3.5 & 4.0 & 4.5 & 5.0 & 5.5 & 6.0 & 6.5 & 7.0 \\

\midrule

\var{front}
& 3.89 & 3.51 & 3.21 & 2.95 & 2.77 & 2.62 & 2.48 & 2.38 & 2.30 & 2.25 \\

\var{back}
& 10.10 & 8.87 & 7.88 & 7.12 & 6.51 & 6.01 & 5.60 & 5.25 & 4.96 & 4.69 \\

\bottomrule
\end{tabular}
}
\end{table}

\subsection{Front-Back Compression}\label{sec:front_back}

The net effect of Formula (\ref{eq:expectation})
is that $P$ has a low entropy.
We report the 0-th order empirical entropy
of $P$ in Table~\ref{tab:entropy}, for different values of $c$.
Specifically, we compare the entropy of $P$ when it stores
the pilots
determined by our approach and when it stores the displacements
following the original FCH procedure.
As expected, the entropy of the pilots
is smaller than that of the displacements,
and actually becomes \emph{much} smaller for increasing $c$.
This clearly suggests that the output of
the search can be compressed very well.
We now make one step further.

We already noticed that the average number of trials
is particularly small
for the first processed buckets because of the
low load factor. This is graphically evident
from Figure~\ref{fig:trials} looking at the first, say, 30\%
of the buckets.
In other words, the first processed buckets
have small pilots.
Now we argue that such buckets
are those corresponding to the first
$p_2=0.3m$ entries of $P$ and hold $p1=0.6n$ keys.
This is a direct consequence of the skewed distribution
of keys into buckets and the order of processing the buckets.
Therefore, the \var{front} part of $P$, $P[0..p_2-1]$,
has a lower entropy compared to its \var{back} part,
$P[p_2..m-1]$.

Table~\ref{tab:entropy_front_back} shows the
entropy of the arrays \var{front} and \var{back}
by varying $c$ and for $n=10^6$.
As evident, the entropy of the \var{front} part
is much smaller than that of the \var{back} part
(by more than $2\times$ on average).
This has two immediate consequences.
The first, now obvious, is that the array \var{front}
is more compressible than \var{back},
and this saves space compared to the case where
$P$ is not partitioned.
Note that this partitioning strategy is guaranteed to
improve compression by virtue of the skewed distribution,
and it is different than partitioning $P$ arbitrarily.
The second, even more important, is that we can use
two different encoding schemes for \var{front} and \var{back}
to help maintaining good lookup performance
and save space.
In fact, since \var{front} holds 60\% of the keys
and its size is 30\% of $|P|$,
it is convenient to use a more time-efficient encoding for \var{front},
which is also more compressible,
and a more space-efficient encoding for \var{back}.

With $P$ rendered as the two partitions \var{front} and \var{back},
evaluating $f(x)$ is still simple.


\vspace{-3mm}
\SetArgSty{textnormal}
\begin{algorithm}
\SetKwBlock{Begin}{}{}
\Begin({$f$}\text{(}$x$\text{)} :)
{
	$i = \func{bucket}(x)$ \\
	\code{if} $i < p_2$ \code{then} $k_i = \var{front}[i]$ \\
	\code{else} $k_i = \var{back}[i-p_2]$ \\
	\code{return} $\func{position}(x,k_i)$
}
\end{algorithm}
\vspace{-3mm}

We then expect the evaluation time of $f(x)$ to be
approximately $(p_1/n)t_f + (1-p_1/n)t_b$
if $x$ is chosen at random from $S$,
with $t_f$ and $t_b$ being the access time of
\var{front} and \var{back} respectively.
In conclusion, because of the values assigned to $p_1$ and $p_2$,
this partitioning strategy allows us to achieve
a trade-off between space and lookup time
depending on what pair of compressors is used
to encode \var{front} and \var{back}.
We will explore this trade-off in Section~\ref{sec:experiments}.

\subsection{Encoding}\label{sec:encoding}

Now that we have achieved the first of the two goals
mentioned at the beginning of Section~\ref{sec:main},
i.e., designing an algorithm that guarantees a compressible output,
we turn our attention to the second one:
devise an encoding
scheme that not only takes advantage of the low entropy of $P$
but also maintains noticeable lookup performance.

For simplicity of exposition, let us now focus
on the case where $P$ is not partitioned
into its \var{front} and \var{back} parts
(the generalization is straightforward).
Since $P$ has a low entropy,
we argue that a \emph{dictionary}-based encoding
is a good match for our purpose:
we collect the \emph{distinct} values of $P$ into an array
$D$, the dictionary,
and represent $P$ as an array of references to $D$'s entries.
Let $r$ be the size of $D$.
As $r$ is smaller than or equal to the number of buckets $m$,
which in turn is smaller than the number of keys $n$,
we can represent each entry of $P$ using $\lceil\log_2 r\rceil$ bits instead
of the $(\lceil\log_2 n\rceil+1)$ bits used by FCH.
The total space usage is given by the
encoded $P$, taking $m\lceil\log_2 r\rceil$ bits,
plus the space of the dictionary $D$.
The latter cost is small compared to that of $P$.
In particular, the larger $c$ is, the smaller
this cost becomes.

The encoding time is linear in the size of $P$, i.e.,
$\Theta(m)$, thus it takes a small fraction
of the total construction.
In particular, all encoding methods
we consider in Section~\ref{sec:experiments} take linear time.

Using the dictionary, the algorithm for $f(x)$ is as follows.


\SetArgSty{textnormal}
\begin{algorithm}
\SetKwBlock{Begin}{}{}
\Begin({$f$}\text{(}$x$\text{)} :)
{
	$i = \func{bucket}(x)$ \\
	$k_i = D[P[i]]$ \\
	\code{return} $\func{position}(x,k_i)$
}
\end{algorithm}

Compared to FCH,
note that we are performing two memory accesses
(for $P$ and $D$) instead of one.
However, since $D$ is small,
its access is likely to be directed to the cache memory
of the target machine,
e.g., L2 or even L1. Therefore, the indirection only
slightly affects lookup performance, as we are going to
show in Section~\ref{sec:experiments}.


Generalizing the approach to the \var{front}
and \var{back} parts of $P$ is straightforward:
each part has its own dictionary and each access
to those arrays is executed as shown in step 3
of the pseudocode.

Other more sophisticated
options are possible to compress $P$,
e.g., Elias-Fano~\cite{Elias74,Fano71},
or the Simple Dense Coding (SDC) by~\citet*{fredriksson2007simple}.
As we are going to see in Section~\ref{sec:experiments},
using these mechanisms is expected to provide superior compression
effectiveness at the price of a slower lookup time.



So far we have explored the effects of the search strategy
on compression effectiveness because,
as introduced at
the beginning of Section~\ref{sec:main},
a compressible output enables the use of a larger $c$ value that
speed up the construction.
Let us recall the concrete example made at the end of Section~\ref{sec:FCH}:
for $n=10^8$ 64-bit keys and $c=3$,
FCH finds a MPHF in 1 hour and 10 minutes
(refer to Section~\ref{sec:experiments}
for a description of our experimental setup).
{\pth} with $c=5.6$
and the introduced front-back encoding
finds a MPHF consuming the same amount of space,
i.e., 3.0 {\bpk}.
However, it does so in 70 seconds,
i.e., $60\times$ faster than FCH.
We will present larger and more detailed experiments
in Section~\ref{sec:experiments}.

\subsection{Limiting the Load Factor}\label{sec:load_factor}

Now, we take a deeper look at searching time.
According to Formula (\ref{eq:expectation}), the searching time
is skewed towards the \emph{end} of the search,
because the expected number of trials rapidly grows
as $\alpha \rightarrow 1$.
This phenomenon is even more evident when using large values of $c$
because a large $c$ lowers the expected number of trials for most of the buckets
by lowering the exponent $|B_i|$ in Formula (\ref{eq:expectation}).
In particular, for larger $c$, a small fraction of time is spent
on most of the keys, and a large fraction is
spent on the last buckets containing
only few keys --- the heavy ``tail'' of the distribution.

\begin{figure}
\centering
\includegraphics[scale=0.8]{{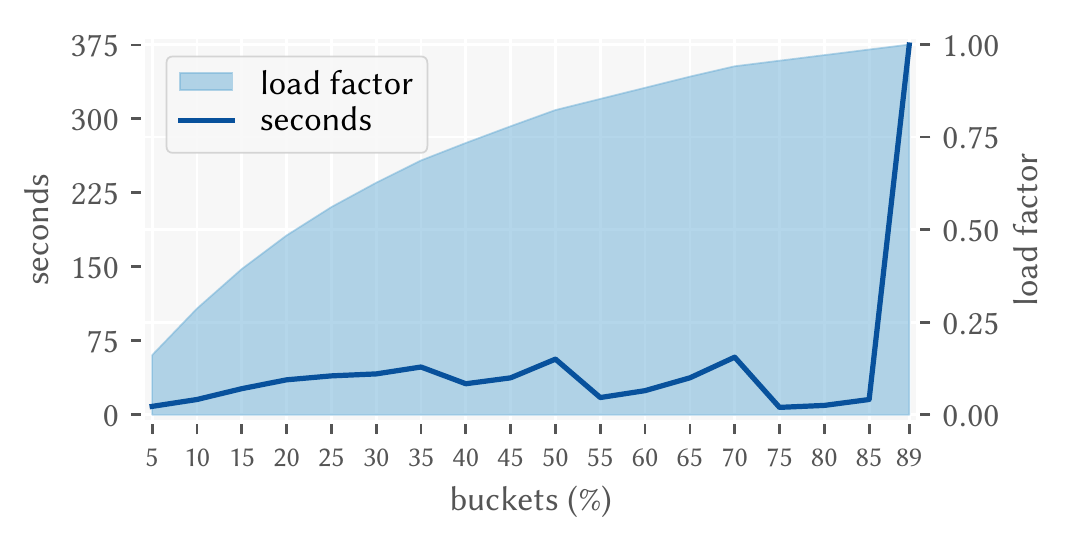}}
\mycaption{Search time spent as buckets are processed
for $n=10^9$ keys and $c=9$. The last 11\% of the buckets are empty.
\label{fig:time_vs_alpha}}
\end{figure}

Figure~\ref{fig:time_vs_alpha} shows an example
of such distribution for $n=10^9$ keys and $c=9$.
Note the high skewness towards the end,
after 85\% of the processed buckets:
more than 40\% of the total search time is spent
for only 1.4\% of the keys falling into the
last 5\% of the non-empty buckets.

To avoid the burden of the heavy tail,
we search for $f$ in a larger space, say $[n^{\prime}=n/\alpha]$
for a chosen maximum load factor $0 < \alpha \leq 1$.
For example, if $\alpha=0.99$, then 1\% extra space
is used to search for $f$.
Limiting the maximum achievable load factor
clearly lowers searching time,
as well as it affects compression effectiveness,
by noting that $\mathbb{E}[k_i]$ \emph{decreases}
as per Formula (\ref{eq:expectation}).
In fact, considering a generic $0 < \alpha \leq 1$,
we obtain the following more general result,
whose proof is omitted due to space constraints.

\textcolor{black}{
\begin{thm}\label{thm:expected_runtime_general}
The expected time of the search,
for $n$ keys and parameters $c > \log_2 e$
and $0 < \alpha \leq 1$,
is $\O\big(n^{ 1 + \Theta(\alpha/c)}\big).$
\end{thm}
%
}

Now, the issue with searching in a space of size
$n^{\prime}=n/\alpha > n$
is that the output of $f$ must be guaranteed to be
minimal, i.e., a value in $[n]$, not in $[n^{\prime}]$.
One can view the strategy as $f$ leaving some ``holes''
in its codomain $[n]$ that must then be filled
in some appropriate manner.
We proceed as follows.
Suppose $L$ is the list of holes \emph{up to} position $n-1$.
There are $|L|$ keys that are actually mapped to positions $p_i \geq n$,
that can fill such holes.
Therefore we materialize an array $\var{free}[0..n^\prime - n - 1]$,
where
$$\var{free}[p_i - n] = L[i], \text{ for each } i=0,\ldots,|L|-1.$$
Note that the space for the array \var{free} is that of
a sorted integer sequence of size $n^\prime - n$,
whose maximum value is less than $n$.
Thus it takes small space, especially if compressed with Elias-Fano, 
i.e., $(n^{\prime}-n)(\lceil\log_2\frac{n}{n^{\prime}-n}\rceil +2+o(1))$ bits.

Let us discuss an explanatory example.
Suppose $n=9$ and $n^{\prime}=14$.
There are $n^{\prime}-n = 14-9 = 5$ holes,
say in positions $[0,2,8,9,12]$.
$L$ is $[0,2,8]$ (holes up to position $n=9$).
Therefore there will be $|L|=3$ keys that are
mapped out of the range $[0..8]$,
and must be those in positions $[10,11,13]$.
Then we assign
$\var{free}[10-9]$ = $L[0]$,
$\var{free}[11-9]$ = $L[1]$,
and $\var{free}[13-9]$ = $L[2]$,
yielding a final $\var{free}[0..4]$ = $[\ast, 0, 2, \ast, 8]$,
where `$\ast$' indicates an unassigned value.

With the array \var{free},
the algorithm for $f(x)$ is as follows.

\SetArgSty{textnormal}
\begin{algorithm}
\SetKwBlock{Begin}{}{}
\Begin({$f$}\text{(}$x$\text{)} :)
{
	$i = \func{bucket}(x)$ \\
	$k_i = P[i]$ \\
	$p = (h(x,s) \oplus h(k_i,s)) \mymod{n^{\prime}}$ \\
	\code{if} $p < n$ \code{then} \code{return} $p$ \\
	\code{else} \code{return} $\var{free}[p-n]$\\
}
\end{algorithm}

The second branch of the conditional (\code{else})
will be taken with probability $\approx(1-\alpha)$
for random queries.
If $\alpha$ is chosen close to 1.0, e.g., 0.99,
the branch will be highly predictable, hence
barely affecting lookup performance.

Our approach guarantees that each key out
of the codomain $[n]$ is mapped back into position
with a \emph{single} access to an array, \var{free}.
As we are going to show in Section~\ref{sec:experiments},
this is considerably faster than the folklore strategy
of filling the array $\var{free}$ with all available free positions
in $[n^{\prime}]$.
In fact, in that case,
a \emph{successor} query must be issued over $\var{free}$
for every position $p$ returned in step 4 of the pseudocode.

Lastly, the algorithm is directly applicable to any compressed
representation of $P$ that supports random access, e.g.,
the front-back scheme with dictionary-based encoding
we have described in the previous sections.


\newpage
\section{Evaluation}\label{sec:experiments}

In this section we present a comprehensive experimental evaluation
of {\pth}. 
All experiments were carried out on a
server machine equipped
with Intel i9-9900K cores (@3.60 GHz),
64 GB of RAM DDR3 (@2.66 GHz), and running Linux 5 (64 bits).
Each core has two private levels of cache memory:
32 KiB L1 cache (one for instructions and one for data);
256 KiB for L2 cache. A shared L3 cache spans 16,384 KiB.
Both construction and lookup algorithms
were run on a single core of the processor,
with the data residing entirely in internal memory.
The implementation of {\pth}
is written in C++ and available at
\url{https://github.com/jermp/pthash}.
The code was compiled with \textsf{gcc} 9.2.1
with optimizating flags
\texttt{-O3} and \texttt{-march=native}.


Lookup time was measured by looking up every single
key in the input, and taking the average time between 5 runs.
For construction time, we report the average between
3 runs.

We build MPHFs
using random integers as input,
which is common practice for benchmarking
hashing-based algorithms~\cite{esposito2020recsplit,limasset2017fast,fan2014cuckoo,graf2020xor,muller2014retrieval},
given that the nature of the data is completely irrelevant
for the space of the data structures.
In our case we generated $n$ 64-bit integers
uniformly at random in the interval $[0,2^{64})$.
We will also evaluate {\pth} on real-world
string collections to further confirm our results.

\begin{table}[t]
\centering
\mycaption{Construction time, space,
and lookup time of {\pth} on $n=10^9$ random 64-bit keys,
for a range of encoding methods and by varying $c$.
\label{tab:encodings}}



\aftertabspace
\subfloat[\boldsymbol{$\alpha=1.00$}]{
\scalebox{\mytablescale}{
\begin{tabular}{c c ccccccc}
\toprule

  \multirow{2}{*}{$c$}
& \multirow{2}{*}{constr. (secs)}
& \multicolumn{7}{c}{{space (\bpk)}} \\

\cmidrule(lr){3-9}

&
& C
& C-C
& D
& D-D
& D-EF
& EF
& SDC
\\

\cmidrule(lr){1-9}







6  &        2441 &     6.22 &             5.14 &        3.21 &                   2.97 &                   2.33 &        2.21 &  2.22 \\
7  &        1579 &     7.49 &             6.09 &        3.75 &                   3.40 &                   2.52 &        2.39 &  2.31 \\
8  &        1190 &     7.76 &             6.31 &        4.28 &                   3.80 &                   2.67 &        2.57 &  2.39 \\
9  &        1010 &     9.03 &             7.22 &        4.82 &                   4.28 &                   2.94 &        2.77 &  2.47 \\
10 &         941 &    10.37 &             8.26 &        5.35 &                   4.65 &                   3.10 &        2.98 &  2.53 \\
11 &         857 &    11.04 &             8.83 &        5.89 &                   5.12 &                   3.32 &        3.19 &  2.60 \\

\cmidrule(lr){1-9}

\multicolumn{2}{c}{lookup ({\nspk})}
 & 36 & 36 & 49 & 48 & 73 & 96 & 165 \\

\bottomrule
\end{tabular}

}
\label{tab:a1.0}
}

\subfloat[\boldsymbol{$\alpha=0.99$}]{
\scalebox{\mytablescale}{
\begin{tabular}{c c ccccccc}
\toprule

  \multirow{2}{*}{$c$}
& \multirow{2}{*}{constr. (secs)}
& \multicolumn{7}{c}{{space (\bpk)}} \\

\cmidrule(lr){3-9}

&
& C
& C-C
& D
& D-D
& D-EF
& EF
& SDC
\\

\cmidrule(lr){1-9}







6  &           1736 &     2.90 &             2.84 &        2.90 &                   2.78 &                   2.31 &        2.17 &  2.23 \\
7  &           1033 &     3.37 &             3.23 &        3.14 &                   3.00 &                   2.44 &        2.27 &  2.31 \\
8  &            749 &     3.57 &             3.41 &        3.30 &                   3.14 &                   2.54 &        2.39 &  2.41 \\
9  &            592 &     3.70 &             3.52 &        3.40 &                   3.31 &                   2.72 &        2.47 &  2.47 \\
10 &            510 &     4.11 &             3.91 &        3.77 &                   3.57 &                   2.80 &        2.59 &  2.53 \\
11 &            450 &     4.14 &             3.92 &        4.14 &                   3.92 &                   2.97 &        2.67 &  2.59 \\

\cmidrule(lr){1-9}

\multicolumn{2}{c}{lookup ({\nspk})}
 & 37 & 37 & 49 & 49 & 75 & 101 & 170 \\

\bottomrule
\end{tabular}

}
\label{tab:a0.99}
}

\vspace{-0.5cm}
\end{table}

\begin{figure}[t]
\centering

\vspace{-3mm}
\hspace{10mm}\includegraphics[scale=0.8]{{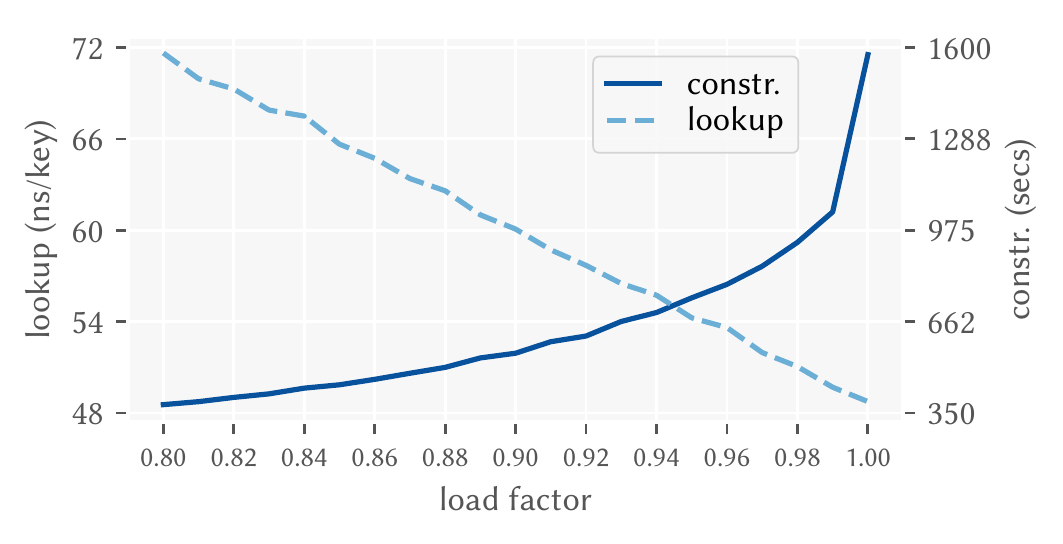}}

\vspace{-3mm}
\hspace{-10mm}\includegraphics[scale=0.8]{{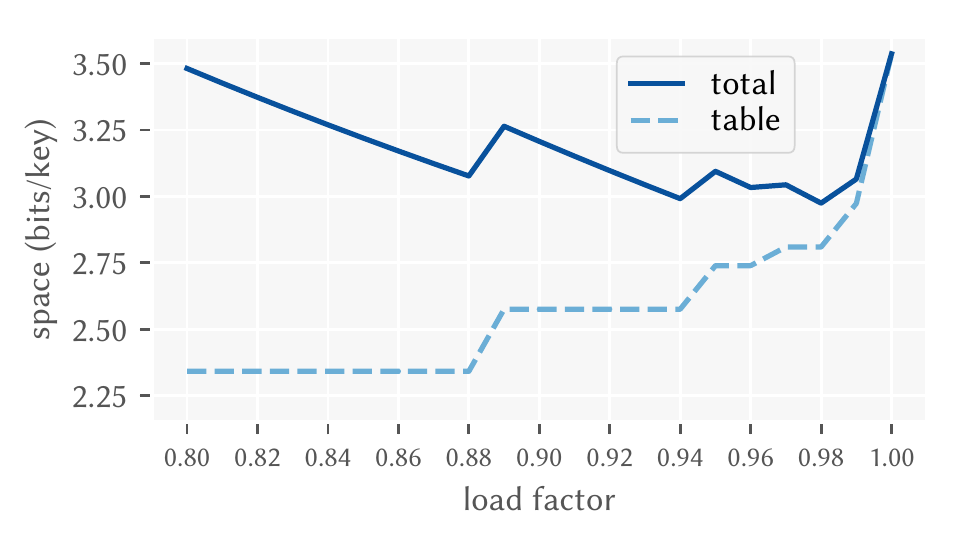}}

\vspace{-5mm}
\mycaption{Trade-off between construction time, lookup time, and space
of {\pth} by varying the load factor,
on $n=10^9$ random 64-bit keys, $c=7$, and the D-D encoding.
\label{fig:construction_vs_lookup}}
\vspace{-6mm}
\end{figure}

\subsection{Tuning}\label{sec:tuning}

In this section, we are interested in tuning {\pth};
we quantify the impact
of (i) different encoding schemes to represent the MPHF
data structure, (ii) front-back compression,
and (iii) varying the load factor.

\parag{Compression Effectiveness}
In Table~\ref{tab:a1.0} we report the performance
of {\pth} with $\alpha=1.0$
in terms of construction time, lookup time,
and {\bpk} rate, for a range of encoding schemes.

We explain the nomenclature adopted to indicate such encodings.
``C'' stands for \emph{compact} and refers to encoding
each value in $P$ with $\lceil\log_2(\max(P)+1)\rceil$ bits
(note that FCH uses this technique, assuming that $\max(P) = n$).
``D'' indicates the \emph{dictionary}-based method described in Section~\ref{sec:encoding}.
``EF'' stands for Elias-Fano~\cite{Elias74,Fano71};
``SDC'' means Simple Dense Coding~\cite{fredriksson2007simple}.
Methods indicated with ``X-Y'' refer to the front-back
compression strategy from Section~\ref{sec:front_back},
where method ``X'' is used for \var{front}
and ``Y'' is used for \var{back}.

Table~\ref{tab:a1.0} is presented to
highlight the spectrum of achievable trade-offs:
from top to bottom, we improve construction time
by increasing $c$;
from left to right we improve space effectiveness
but degrade lookup efficiency.

We recall that construction time includes the time of
mapping, sorting, searching, and encoding.
The most time
consuming step is the search, especially for small $c$.
For the experiment reported in Table~\ref{tab:encodings},
mapping+sorting took 105 seconds, whereas encoding time
is essentially the same for all the different methods tested.
It ranges
from 10 
to 15 
seconds by varying $c$ from 6 to 11,
as $c$ affects the number of buckets used.
All the rest of the time is spent during the searching step.
Furthermore, we make the following observations.


\return\noindent$\bullet$
Front-back compression pays off for the reasons
we explained in Section~\ref{sec:front_back}, as it always
improves space effectiveness, i.e., C-C and D-D are
more compact than C and D respectively,
while preserving their relative lookup efficiency.

\noindent$\bullet$
D and D-D
are always more compact than C and C-C,
and slightly affect lookup time (+12 {\nspk} on average
for $n=10^9$, but smaller for smaller values of $n$),
hence confirming that dictionary-based encoding is a good match
as 
explained
in Section~\ref{sec:encoding}.

\noindent$\bullet$
EF and SDC are better suited for space effectiveness
but also $1.5-3\times$ slower on lookup compared to D-D.

\noindent$\bullet$
The configuration D-EF stands in trade-off
position between D-D and EF, as the use of EF on the \var{back}
part improves space but slows down lookup.


\parag{Varying the Load Factor}
We explained in Section~\ref{sec:load_factor},
that varying $\alpha$
trades off between construction time and lookup efficiency.
Figure~\ref{fig:construction_vs_lookup} shows a pictorial example
of this trade-off
by varying $\alpha$ from 0.80 to 1.00 with step 0.01.


\return\noindent$\bullet$
Construction time decreases significantly,
especially in the range $\alpha \in [0.9,1.0]$.
In particular, note the sharp decrease when passing
from $\alpha=1.0$ to $\alpha=0.99$
as a consequence of avoiding the heavy ``tail''
already observed in Figure~\ref{fig:time_vs_alpha}.

\noindent$\bullet$
Lookup time, instead, increases
as $\alpha$ decreases, but at a much slower pace
thanks to the fast re-ranking of keys
we described in Section~\ref{sec:load_factor}
to guarantee that the output of the function is minimal.
In fact, while we are able to obtain a $2\times$
faster construction when passing from, say, $\alpha=1.0$
to $\alpha=0.94$, we only increase lookup time
by 6 {\nspk}.
At the other edge of the spectrum visible
in Figure~\ref{fig:construction_vs_lookup},
if we use $\alpha=0.80$ we obtain a $3\times$ faster construction
but also pay $\approx$22 {\nspk} more on lookup.

\noindent$\bullet$
The other relevant advantage is that
using a lower load factor
does \emph{not} consume more space but actually even less,
for the reasons explained in Section~\ref{sec:load_factor}.
In Figure~\ref{fig:construction_vs_lookup} we show
the total space of the MPHF and that taken by the
pilots table $P$ alone.
When $\alpha < 1.0$, the total space is given by
the space of $P$ plus that of the \var{free} array
that we compress with Elias-Fano
as explained in Section~\ref{sec:load_factor}.


\begin{table}[t]
\centering
\mycaption{The performance of FCH for $n=10^8$ random 64-bit keys,
and some {\bpk} rates.
For comparison, also the performance of
{\pth} with $\alpha=0.99$ and encoding
C, D-D, and D-EF, is reported.}
\aftertabspace
\scalebox{\mytablescale}{

\begin{tabular}{cc cc@{\hspace{1mm}} cc@{\hspace{1mm}} cc@{\hspace{1mm}} cc}
\toprule

\multicolumn{2}{c}{\multirow{2}{*}{space (\bpk)}} & \multicolumn{7}{c}{constr. (secs)} \\

\cmidrule(lr){3-9}


%
%
%
%

&
& FCH
& \multicolumn{2}{c}{C}
& \multicolumn{2}{c}{D-D}
& \multicolumn{2}{c}{D-EF}
\\

\cmidrule(lr){1-9}

\multicolumn{2}{c}{2.50} & --- & 614 & & 173 & & 44 & \\
\multicolumn{2}{c}{3.00} & 4286 & 62 & \tabtimes{69} & 37 & \tabtimes{116} & 22 & \tabtimes{195} \\
\multicolumn{2}{c}{3.50} &  985 & 29 & \tabtimes{34} & 27 & \tabtimes{36} & 20 & \tabtimes{49} \\
\multicolumn{2}{c}{4.00} &  463 & 25 & \tabtimes{18} & 22 & \tabtimes{21} & 20 & \tabtimes{23} \\
\multicolumn{2}{c}{4.50} &  340 & 22 & \tabtimes{15} & 20 & \tabtimes{17} & 20 & \tabtimes{17} \\
\multicolumn{2}{c}{5.00} &  145 & 21 & \tabtimes{7} & 20 & \tabtimes{7} & 20 & \tabtimes{7} \\
\cmidrule(lr){1-9}

\multicolumn{2}{c}{lookup ({\nspk})}
 & 30 & \multicolumn{2}{c}{28} & \multicolumn{2}{c}{35} & \multicolumn{2}{c}{55} \\

\bottomrule
\end{tabular}

}
\label{tab:search_FCH}
\vspace{-0.3cm}
\end{table}

\return
In Table~\ref{tab:a0.99} we report the result of the
same experiment in Table~\ref{tab:a1.0}
but with load factor $\alpha = 0.99$.
The tables are shown next to each other
to better highlight the comparison.
As already noted in Figure~\ref{fig:construction_vs_lookup},
using just 1\% extra space for the search
already introduces important advantages,
that are observed for \emph{any} encoding method and
\emph{any} value of $c$:
(i) 25--35\% reduced construction time;
(ii) reduced space usage
(with noteworthy improvements for the encodings C and D);
(iii) preserved lookup efficiency.

\parag{Speeding Up the Search}
As a last experiment in this section,
Table~\ref{tab:search_FCH} shows the speed up
factors achieved by some {\pth} configurations
over the construction time of FCH,
for the \emph{same} final {\bpk} rates.
Even using the simple C encoding for {\pth}
yields $7-69\times$ faster construction with
equal (or better) lookup efficiency.
Moving to the right-hand side of the table
brings further advantages in
construction time at the price of a penalty in lookup.

\begin{table}[t]
\centering
\mycaption{Construction time,
space, and lookup time for a range of methods
on 64-bit random keys.}
\aftertabspace
\scalebox{\mytablescale}{
\setlength{\tabcolsep}{1pt}
\begin{tabular}{
l c
ccc c ccc}

\toprule

\multirow{3}{*}{Method}
&& \multicolumn{3}{c}{{$n=10^8$}}
&& \multicolumn{3}{c}{{$n=10^9$}} \\

\cmidrule(lr){3-5}
\cmidrule(lr){7-9}

&& constr. & space & lookup
&& constr. & space & lookup \\

&& \small (secs) & \small ({\bpk}) & \small ({\nspk})
&& \small (secs) & \small ({\bpk}) & \small ({\nspk}) \\

\midrule


FCH, $c=3$
    && {4286} & {3.00} & {30}
    && {---} & {---} & {---} \\
FCH, $c=4$
    && {463} & {4.00} & {30}
    && {15904} & {4.00} & {35} \\
FCH, $c=5$
    && {145} & {5.00} & {30}
    && {2937} & {5.00} & {35} \\
FCH, $c=6$
    && {81} & {6.00} & {30}
    && {2133} & {6.00} & {35} \\
FCH, $c=7$
    && {68} & {7.00} & {30}
    && {1221} & {7.00} & {35} \\

\midrule

CHD, $\lambda=4$
    && {121} & {2.17} & {204}
    && {1972} & {2.17} & {419} \\
CHD, $\lambda=5$
    && {358} & {2.07} & {204}
    && {5964} & {2.07} & {417} \\
CHD, $\lambda=6$
    && {1418} & {2.01} & {197}
    && {23746} & {2.01} & {416} \\

\midrule

EMPHF
    && {24} & {2.61} & {147}
    && {374} & {2.61} & {199} \\


\midrule

GOV
    && 85 & 2.23 & 110
    && 875 & 2.23 & 175 \\

\midrule

BBHash, $\gamma=1$
    && {14} & {3.06} & {119}
    && {253} & {3.06} & {170} \\
BBHash, $\gamma=2$
    && {10} & {3.71} & {108}
    && {152} & {3.71} & {143} \\
BBHash, $\gamma=5$
    && {8} & {6.87} & {98}
    && {100} & {6.87} & {113} \\

\midrule

RecSplit, $\ell$=5, $b$=5
    && {20} & {2.95} & {157}
    && {233} & {2.95} & {220} \\
RecSplit, $\ell$=8, $b$=100
    && {92} & {1.80} & {124}
    && {936} & {1.80} & {204} \\
RecSplit, $\ell$=12, $b$=9
    && {569} & {2.23} & {110}
    && {5700} & {2.23} & {197} \\

\midrule

{\pth} && & & && & & \\

(i) C-C, $\alpha$=0.99, $c$=7
    && 42 & 3.36 & 28
    && 1042 & 3.23 & 37 \\
(ii) D-D, $\alpha$=0.88, $c$=11
    && 19 & 4.05 & 46
    && 308 & 3.94 & 64 \\
(iii) EF, $\alpha$=0.99, $c$=6
    && 45 & 2.26 & 49
    && 1799 & 2.17 & 101 \\


(iv) D-D, $\alpha$=0.94, $c$=7
    && 26 & 3.23 & 37
    && 689 & 2.99 & 55 \\

\bottomrule
\end{tabular}
}
\label{tab:overall}
\end{table}

\subsection{Overall Comparison}\label{sec:overall_comparison}
In this section we compare {\pth} with
the state-of-the-art techniques reviewed in Section~\ref{sec:related_work}.


\return\noindent$\bullet$
FCH~\cite{fox1992faster} 
---
It is the only algorithm that we re-implemented (in C++) faithfully
to the original paper\footnote{The popular CMPH library
contains an implementation of FCH
that we could not use because it is
orders of magnitude
slower than our implementation.
}.
We tested the algorithm with $c=3,\ldots,7$ as to almost
cover the spectrum of {\bpk} rates achieved by the other methods.

\noindent$\bullet$
CHD~\cite{belazzougui2009hash}
---
We tested the method with parameter $\lambda = 4, 5, 6$.
We were unable to use $\lambda=7$
for more than a few thousand keys, as already
noted in prior work~\cite{esposito2020recsplit}.

\noindent$\bullet$
EMPHF~\cite{belazzougui2014cache}
---
It is an efficient implementation of the method
based on peeling random hypergraphs. Although the library
can also target external memory, we run the algorithm
in internal memory.

\noindent$\bullet$
GOV~\cite{genuzio2016fast,genuzio2020fast}
---
It is a method based on solving random linear systems
via the Gaussian elimination technique.

\noindent$\bullet$
BBHash~\cite{limasset2017fast}
---
It is tested with parameter $\gamma = 1,2,5$
as suggested in the original paper.
The construction can be multi-threaded, but we used one
single thread as to ensure a fair comparison.

\noindent$\bullet$
RecSplit~\cite{esposito2020recsplit}
---
We tested the method using the same configurations used
by the authors in their paper, as to offer different trade-offs
between construction time and space effectiveness.


\return
For all the above methods we used the source code provided
by the original authors (see the References for the corresponding
GitHub repositories) and set up a benchmark
available at

\url{https://github.com/roberto-trani/mphf_benchmark}.

\noindent
All implementations are in C/C++
except for GOV whose construction is only available in Java.

The results in Table~\ref{tab:overall}
are strongly consistent with those reported
in recent previous work~\cite{limasset2017fast,esposito2020recsplit}.
Out of the many possible configurations for {\pth},
we isolate the following four ones.


\return\noindent(i)
\emph{Optimizing lookup time} ---
C-C encoding, $\alpha=0.99$, $c=7$.
{\pth} in this configuration achieves the same lookup time
as FCH but in much better compressed space.
It is similar in space to BBHash with $\gamma=1,2$
but $3-4.5\times$ faster at lookup.
Compared to other more space-efficient methods,
{\pth} is $0.5-1$ bit/key larger but also $5.4-11\times$
faster at lookup.

\noindent(ii)
\emph{Optimizing construction time} ---
D-D encoding, $\alpha=0.88$, $c=11$.
This configuration shows that {\pth}
is competitive in construction time with most of
the other techniques,
at the price of a larger space consumption.
In any case, lookup performance
is significantly better, by at least a factor of $2\times$.

\noindent(iii)
\emph{Optimizing space effectiveness} ---
EF encoding, $\alpha=0.99$, $c=6$.
In this configuration {\pth} achieves a space effectiveness
comparable with that of the most succinct methods,
i.e., CHD and RecSplit, while still being $2-4\times$ faster
at lookup than those methods.

\noindent(iv)
\emph{Optimizing the general trade-off} ---
D-D encoding, $\alpha=0.94$, $c=7$.
This configuration tries to achieve a balance between
the other three configurations,
combining good space effectiveness and construction time,
with very fast lookup evaluation.


\return
The evident takeaway message emerging from the comparison
in Table~\ref{tab:overall}
is that \emph{there is a configuration of {\pth} that takes
space similar to another method but provides remarkably
better lookup performance, with
feasible or better construction speed.}

\subsection{Performance on Variable-Length Keys}\label{sec:strings}
In this section, we evaluate {\pth} on real-world
datasets of variable-length keys.
We used natural-language $q$-grams
as they are in widespread use in IR, NLP, and machine-learning
applications;
URLs are interesting as they represent a sort of ``worst-case'' input
given their very long average length.
More specifically, we used the 1-grams and 2-grams from the English GoogleBook (version 2)
corpus\footnote{\url{http://storage.googleapis.com/books/ngrams/books/datasetsv2.html}},
that are \num{24357349} and \num{665752080} in number,
respectively.
For URLs, we used those of the $\approx$50 million Web pages
in the ClueWeb09 (Category B) dataset\footnote{\url{https://lemurproject.org/clueweb09}},
and those collected in 2005
from the UbiCrawler~\cite{boldi2004ubicrawler}
relative to the .uk domain\footnote{\url{http://data.law.di.unimi.it/webdata/uk-2005/uk-2005.urls.gz}},
for a total of
\num{49937704} and \num{39459925} URLs respectively.

While the space of the MPHF data structure is independent
from the nature of the data,
we choose datasets of increasing average key size
to highlight the difference in construction and lookup
time compared to fixed-size 64-bit keys.
Recall that {\pth} hashes each input key only once
during construction (to distribute keys into buckets),
and once per lookup.
Thus, we expect the timings to increase \emph{by a constant}
amount per key, i.e.,
by the difference between the time to hash
a long key and a 64-bit key.

\begin{table}[t]
\centering
\mycaption{Construction time,
space, and lookup time of {\pth} on some string collections.
For comparison, also the performance on 64-bit (fixed-length)
random keys is reported.
The used {\pth} configuration is (iv) --- D-D, $\alpha=0.94$, $c=7$.}
\aftertabspace
\scalebox{\mytablescale}{


\begin{tabular}{l cccc}

\toprule

\multirow{2}{*}{Collection} & avg. key size & constr. & space & lookup \\

& \small (bits) & \small (secs) & \small ({\bpk}) & \small ({\nspk}) \\

\midrule

GoogleBooks 1-grams
& \,\,82.32 & 5 & 3.18 & 22 \\

64-bit keys
& \,\,64.00 & 5 & 3.18 & 16 \\

\midrule

GoogleBooks 2-grams
& 137.68 & 430 & 2.97 & 63 \\

64-bit keys
& \,\,64.00 & 428 & 2.97 & 54 \\

\midrule

ClueWeb09 URLs
& 437.76 & 12 & 3.07 & 49 \\

64-bit keys
& \,\,64.00 & 11 & 3.07 & 24 \\

\midrule

UK2005 URLs
& 570.96 & 9 & 3.11 & 48 \\

64-bit keys
& \,\,64.00 & 9 & 3.11 & 22 \\

\bottomrule
\end{tabular}
}
\label{tab:perf_variable_length_keys}
\vspace{-0.5cm}
\end{table}

The performance of {\pth} on such variable-length keys
is reported in Table~\ref{tab:perf_variable_length_keys},
under the configuration (iv) --- D-D, $\alpha=0.94$, $c=7$.
Space effectiveness does not change as expected
between real-world datasets and random keys.
Construction time does not change either,
because of the difference in scale between a process
that takes seconds and a hash calculation taking nanoseconds:
a constant amount of nanoseconds per key
during \emph{only} the mapping step does not impact.
Instead, lookup time grows proportionally to the length of the keys
showing that the hash calculation contributes to most
of the lookup time of {\pth}.
More specifically, it increases by
$6-9$ {\nspk} on the $q$-gram datasets, for $1.3-2\times$ longer keys;
and by $25-26$ {\nspk} on URLs,
for $7-9\times$ longer keys.
These absolute increments show the impact of the hashing of longer keys in a way
that is independent of the encoding scheme and size of the dataset.

Concerning the other
methods tested in Section~\ref{sec:overall_comparison},
a similar increase was observed for those
hashing the key once per lookup (like {\pth}),
or much worse for those that hash the key
several times per lookup, e.g., BBHash.



\section{Conclusions}\label{sec:conclusions}
We presented {\pth}, an algorithm that builds
minimal perfect hash functions in compact space
and retains excellent lookup performance.
The result was achieved via a careful revisitation
of the framework introduced by~\citet*{fox1992faster} (FCH)
in 1992.

We conduct a comprehensive experimental evaluation and show that
{\pth} takes essentially the same space as that of previous
state-of-the art algorithms
but provides $2-4\times$ better lookup time.
While space effectiveness remains a very important
aspect, efficient lookup time is even more important
for the minimum perfect hashing problem
and its applications.
Our C++ implementation is publicly available
to encourage the use of {\pth} and
spur further research on the problem.

Future work will target parallel and external-memory construction,
e.g., by splitting the input into chunks
and building an independent MPHF on each chunk~\cite{botelho2013practical};
and devise even more succinct encodings.
It would be also interesting to generalize the algorithm
to build other types of functions,
such as perfect (non-minimal), and
$k$-perfect hash functions.

\begin{acks}
This work was partially supported by the projects: MobiDataLab (EU H2020 RIA, grant agreement N\textsuperscript{\b{o}}101006879) and OK-INSAID (MIUR-PON 2018, grant agreement N\textsuperscript{\b{o}}ARS01\_00917).
\end{acks}

\balance
\renewcommand{\bibsep}{3.0pt}
\bibliographystyle{ACM-Reference-Format}
\bibliography{biblio}

\end{document}